\documentclass{emulateapj}

\usepackage{times}
\usepackage{mathptm}

\renewcommand{\thefootnote}{\fnsymbol{footnote}}

\slugcomment{}
\shorttitle{Environmental Effects on Virgo Spirals}
\shortauthors{Nakanishi et al.}

\begin{document}
\title{Environmental Effects on Gaseous Disks of the Virgo Spiral Galaxies\footnotemark[1]}
\author{Hiroyuki Nakanishi\altaffilmark{1,2}, Nario Kuno\altaffilmark{2,3}, Yoshiaki Sofue\altaffilmark{1}, Naoko Sato\altaffilmark{4,5}, Naomasa Nakai\altaffilmark{6}, Yasuhiro Shioya\altaffilmark{7,8}, Tomoka Tosaki\altaffilmark{9,2}, Sachiko Onodera\altaffilmark{1}, Kazuo Sorai\altaffilmark{4}, Fumi Egusa\altaffilmark{1},  and Akihiko Hirota\altaffilmark{10,2}}
\email{hnakanis@nro.nao.ac.jp}

\altaffiltext{1}{Institute of Astronomy, The University of Tokyo, 2-21-1 Osawa, Mitaka, Tokyo 181-0015, Japan.}
\altaffiltext{2}{Nobeyama Radio Observatory, Minamimaki, Minamisaku, Nagano 384-1305, Japan.}
\altaffiltext{3}{The Graduate University for Advanced Studies (SOKENDAI), 2-21-1 Osawa, Mitaka, Tokyo 181-0015, Japan.}
\altaffiltext{4}{Division of Physics, Graduate School of Science, Hokkaido University, Sapporo 060-0810, Japan.}
\altaffiltext{5}{Center for Education and Research of Lifelong Learning, Wakayama university, Wakayama 640-8510, Japan.}
\altaffiltext{6}{Institute of Physics, University of Tsukuba, Ten-nodai, 1-1-1 Tsukuba, Ibaraki 305-8577, Japan.}
\altaffiltext{7}{Astronomical Institute, Graduate School of Science, Tohoku University, Aramaki, Aoba, Sendai 980-8578, Japan.}
\altaffiltext{7}{Department of Physics, Faculty of Science, Ehime University, Matsuyama 790-8577, Japan}
\altaffiltext{9}{Gunma Astronomical Observatory, 6860-86 Nakayama, Takayama, Agatsuma, Gunma 377-0702, Japan.}
\altaffiltext{10}{Department of Astronomy, School of Science, The University of Tokyo, Bunkyo-ku, Tokyo 113-0033, Japan.}

\footnotetext[1]{A part of this work was carried out under the common use observation program at the Nobeyama Radio Observatory (NRO). NRO is a branch of the National Astronomical Observatory.}
\begin{abstract}
We found high molecular fractions ($f_{\rm mol}$; ratio of the molecular to total gas surface densities) in three of five Virgo spiral galaxies in spite of their low total gas column density, based on $^{12}$CO$(J=1-0)$ observations with the Nobeyama 45 m telescope equipped with a multi-beam receiver, BEARS. We interpret this as a result of environmental effects.
Combining the CO data with HI data, the relationship between the surface density of the total gas (HI plus H$_2$) and $f_{\rm mol}$ indicates that the three galaxies near the cluster center have larger $f_{\rm mol}$ values than expected for field galaxies, while the others show normal $f_{\rm mol}$. The large $f_{\rm mol}$ is interpreted as being due either to effective HI gas stripping, even in the inner disks, or to large ISM pressure induced by the high ICM pressure and/or ram pressure, although the possibility of an unusually high metallicity cannot be ruled out.

\end{abstract}
\keywords{galaxies: clusters: individual (Virgo),  galaxies: ISM,  ISM: molecules,  radio lines: ISM}

\section{Introduction}
Galaxy members in a cluster of galaxies are embedded in a hot diffuse intra-cluster medium (ICM) emitting X-rays, and suffering various environmental effects. 
Since Virgo is one of the nearest clusters, we can study the environmental effects there in greatest detail. 

Observation in the 21 cm line show that the HI gas is deficient for galaxies near the center of the Virgo Cluster (e.g., Cayatte et al. 1990). 
This is interpreted as being a result of ram pressure stripping.  
\citet{vol01} simulated the ram pressure effect by a numerical calculation and matched well the observed distortion of the HI disk due to the ram pressure. \citet{ken88, ken89} observed the Virgo spiral galaxies in the CO$(J=$1--0) line using the 14 m telescope of the Five College Radio Astronomy Observatory (FCRAO) and found that the molecular gas content appears to be normal, even in HI deficient galaxies. They reported that HI-deficient galaxies have a high ratio of CO flux to HI flux.

{Whether a galaxy suffers environmental effects or not, the inner disk of a spiral galaxy is dominated by molecular gas, because in general the inner disk is originally dominated by molecular gas. 
Therefore, it has not been clarified whether the inner disk is affected by environmental effects or not. 

We can enumerate three possibilities for the physical condition of the inner disk: (1) ram pressure stripping occurs only in the outer HI disk, and the inner disk suffers no environmental effect, (2) ram pressure stripping occurs in the inner disk as well as the outer disk, but only HI gas is selectively stripped, (3) the ISM pressure of the inner disk increases due to external pressures, and molecular gas formation is enhanced. The molecular gas can be dominant in the inner disk in any of these three possibilities.

In order to develop our understanding of the physical condition of the gas in inner disks of cluster spiral galaxies, we must pay attention to the molecular fraction ($f_{\rm mol}$; ratio of molecular to total gas surface density). In the first possibility, there would be no difference in the molecular fraction, $f_{\rm mol}$, in the inner disk between cluster and field galaxies. However, the $f_{\rm mol}$ of cluster galaxies would be larger than that of field galaxies in the second and third possibilities.

In this paper, we show results of the highest resolution single-dish observations in the CO line of five Virgo Cluster galaxies achieved with the Nobeyama 45 m telescope equipped with a multi-beam receiver, BEARS. We also compare the CO data with HI data with almost the same resolution obtained with the Very Large Array (VLA) C and D-configurations. These data sets enable us to investigate the molecular fraction at each point of the spiral galaxies with a fine scale, while earlier research has dealt with the total amounts of HI and CO (H$_2$) gases in galaxies. Based on clues from $f_{\rm mol}$, we quantitatively discuss the physical condition of the gas disks in terms of (1) HI stripping, (2) external pressure due to the ICM or ram pressure, (3) intrinsic metallicity in a galaxy, and (4) the UV field of a galaxy. }
The distance of the Virgo cluster is taken to be 16.1 Mpc \citep{fer96}.

\section{Observations and Data}
The $^{12}$CO$(J=1-0)$ observations of the Virgo Cluster spirals were made from 2002 December to 2004 April with the Nobeyama 45 m radio telescope. The half-power beam-width was $15\arcsec$ (1.2 kpc at the Virgo Cluster) at 115 GHz. We used a focal plane array receiver, BEARS (SIS 25-BEam Array Receiver System), which consisted-of 25 beams ($5\times 5$) and was operated in the double-side-band (DSB) mode \citep{sun00}. The separation between the beams was $41\farcs1$. 
Twenty-five digital auto-correlators with a 500 kHz resolution (1.3 km s$^{-1}$ at 115 GHz) and a 512 MHz coverage (1332 km s$^{-1}$) \citep{sor00a} were used as spectrometers.  

Calibration of the line intensity was made with an absorbing chopper wheel in front of the receiver, which yielded the antenna temperature $T_{\rm A}^*$ corrected for the atmospheric and antenna ohmic losses. Since the scale of the intensity varies with the beams, mainly due to the difference of the side band ratio between the upper side band (USB) and the lower side band (LSB), we scaled the observed DSB intensity to the SSB (single side band) intensity using scaling factors that were determined by comparing the CO intensities of NGC 7538 measured with BEARS and a single-beam SSB receiver, SIS100. The main beam efficiency of the telescope was $\eta_{\rm MB} = 0.40$ at 115 GHz, and the intensity given in this paper is the main beam brightness temperature, defined by $T_{\rm MB} \equiv T_{\rm A}^*/\eta_{\rm MB}$. 

We mapped five Virgo galaxies with a grid spacing of $10\farcs3$. The number of total observed points for each galaxy was 576 and the mapped area was $3\farcm95 \times 3\farcm95$, except for NGC 4254 and NGC 4569, whose number of observed points was 1056 and the area was $3\farcm95 \times 7\farcm38$. 
The observations were made in the position-switching mode with an off-position at an offset of $7\arcmin$ from the centers of the galaxies in azimuth. The typical rms noise of the main beam brightness temperature was $\Delta T_{\rm MB} = 0.09 - 0.18$ K per velocity channel. 
Pointing of the antenna was calibrated by observing the continuum point source 3C273 at 43 GHz, and its typical error was less than $5\arcsec$ (peak value).  Our observed samples were selected out of the galaxies in \citet{sof03}, and the observing parameters are listed in Table 1.

\vskip 3mm
\centerline{--- Table 1   ---}
\vskip 3mm

To compare with the distribution of CO, we used the HI data of the target galaxies mapped with the VLA. The HI data were adopted from \citet{pho93} for NGC 4254, from \citet{cay90} for NGC 4402, NGC 4569, and NGC 4579, and from \citet{pho95} for NGC 4654. The spatial resolution of the HI maps was about $20\arcsec$, which is close to that of our CO maps. Table 2 summarizes the parameters of the HI data.  

\vskip 3mm
\centerline{--- Table 2   ---}
\vskip 3mm

Figure 1 shows the distribution of the CO integrated intensity, $I_{\rm CO} \equiv \int T_{\rm MB} dv$ (K km s$^{-1}$), ({\it color}) and the HI intensity ({\it contours}) of the five galaxies. 
The CO emissions distribute in the inner disks, and the extensions are typically up to radii of $\sim 1\farcm5$ (7 kpc). 

The HI gas in NGC 4254 and NGC 4654 is abundant and its distribution is strongly distorted. The HI disks of NGC 4402 and NGC 4569 are truncated at the edges of the CO disks. In NGC 4579, the HI gas surrounds the central CO disk where the HI gas is deficient.  

\vskip 3mm
\centerline{--- Figure 1   ---}
\vskip 3mm

\section{Environmental Effects on Gaseous Disks}
\subsection{Molecular Fraction as a Function of the Surface Density of the Total Gas}
The fraction of molecular gas to the total gas (HI gas plus H$_2$ gas), $f_{\rm mol}$, is determined by the ISM (interstellar medium) pressure $P$, metallicity $Z$, and UV radiation $U$ \citep{elm93}. Here the molecular fraction $f_{\rm mol}$ is defined as
\begin{equation}
f_{\rm mol} \equiv {\Sigma_{{\rm H}_2} \over \Sigma_{\rm  HI}+\Sigma_{{\rm  H}_2}}, 
\end{equation}
with $\Sigma_{{\rm  HI}}$ and $\Sigma_{{\rm  H}_2}$ being the surface mass densities of HI and H$_2$, respectively. Since the ISM pressure is approximately proportional to the square of the gas surface density \citep{elm93}, 
\begin{equation}
\label{eq-pressure}
{P\over P_0} = {\Sigma^2 \over \Sigma^2_0}={(\Sigma_{{\rm  HI}}+\Sigma_{{\rm  H}_2})^2 \over (\Sigma_{{\rm HI}}+\Sigma_{{\rm H}_2})^2_0}, 
\end{equation}
$f_{\rm mol}$ is expressed as a function of $\Sigma$, when $Z$ and $U$ are given. 
\citet{hon95}, \citet{kun95}, and \citet{sor00b} investigated $f_{\rm mol}$ in nearby spiral galaxies, and showed that it is well reproduced by a model of \citet{elm93}. 
Figure 2 shows model curves of the $\Sigma$--$f_{\rm mol}$ relation calculated based on \citet{elm93}, where we scale $P$, $U$, and $Z$ with the values of the solar neighborhood: $P_0$, $U_0$, and $Z_0$. The gas surface density at the solar neighborhood, $\Sigma_0$, is taken to be 8 $M_\odot$ pc$^{-2}$ \citep{san84}, which is adopted to be proportional to $P_0^{1/2}$.  We show model curves taking UV radiations of $0.1 U_0$, $1 U_0$, and $10 U_0$ and metallicities of $0.1 Z_0$, $1 Z_0$, and $10 Z_0$.

\vskip 3mm
\centerline{--- Figure 2   ---}
\vskip 3mm

Figure 3 shows $f_{\rm mol}$ against $\Sigma$ for each observed point of the five galaxies, where the CO-to-H$_2$ conversion factor was adopted to be $X_{\rm CO} = 1.0\times10^{20}$ H$_2$ cm$^{-2}$ K$^{-1}$ km$^{-1}$ s \citep{nak95}. The dashed curves in all panels denote the averaged $f_{\rm mol}$ of NGC 4254 and NGC 4654. The molecular fractions, $f_{\rm mol}$, of NGC 4254 and NGC 4654 show similar curves to the models with $U=1U_0$ and $Z=1Z_0$, which is expected for field galaxies. 

On the contrary, the other three galaxies (NGC 4402, NGC 4569, and NGC 4579) present larger molecular fractions than the dashed curve. NGC 4402 shows an extraordinarily large $f_{\rm mol}$ in spite of the small surface density of the gas. For NGC 4569 the plotted points are scattered on the upper side of the dashed curve. NGC 4579 shows a very large $f_{\rm mol}$ in spite of the small surface density of the gas as NGC 4402. In the three galaxies, $f_{\rm mol}$ is always larger than the dashed curve.

\vskip 3mm
\centerline{--- Figure 3   ---}
\vskip 3mm

\subsection{Origin of the Unusual Molecular Fraction in the Virgo Spirals}
\subsubsection{Ram pressure Stripping}
Ram pressure stripping is a possible interpretation of the unusually large $f_{\rm mol}$ of NGC 4402, NGC 4569, and NGC 4579. 

Ram pressure stripping would occur if the criterion  
\begin{equation}
\left({\rho_{\rm ICM}\over\rho_{\rm ISM}}\right) \left({R\over d}\right)\left({\delta v \over V_{\rm rot}}\right)^2> 1
\end{equation} 
is satisfied, where $\rho_{\rm ISM}$ and $\rho_{\rm ICM}$ are the volume number density of the ISM and the ICM, respectively, $R$ is the galactocentric radius of the element, $d$ is the thickness of the gas disk, $\delta v$ is the relative velocity of an ISM cloud against the ICM, and $V_{\rm rot}$ is the rotational velocity of the galaxy. 

For a simple case, we consider a gas disk with thickness $d\sim 0.05$ kpc, rotating at $V_{\rm rot} \sim 200$ km s$^{-1}$. 
For a typical ICM wind with $\rho_{\rm ICM} \sim 10^{-4}$ H cm$^{-1}$, and $\delta v \sim 1000$ km s$^{-1}$ in the Virgo Cluster \citep{hid02}, we obtain for the criterion of where ram pressure stripping can occur, 
\begin{equation}
\label{criterion}
\rho_{\rm ISM} < 0.05 \left( {R\over {\rm kpc}} \right) [\mbox{H cm}^{-3}]. 
\end{equation}

The HI gas can be dominant in the inter-arm region even in inner disks, and its density typically ranges from $0.01$ to $1$ H cm$^{-3}$. Hence, the HI gas can be stripped even in the inner disks. On the other hand, the molecular cloud can hardly be stripped because the gas density in the molecular cloud is $ 10 - 1000$ H$_2$ cm$^{-3}$.    
Therefore, the HI gas can be stripped selectively from a disk, although the molecular gas still remains in the disk. 
In this case, $f_{\rm mol}$ in the $\Sigma$---$f_{\rm mol}$ diagram moves toward the upper-left side of the $\Sigma$---$f_{\rm mol}$ curve of field galaxies.

NGC 4402, NGC 4569, and NGC 4579 are located within the projected radius of $2\arcdeg$ from M87 at the center of the cluster, and are known to be HI deficient. Hence, they must have experienced ram pressure stripping of the HI gas. 
We emphasize that in these three galaxies, (1) the HI disks are restricted within the central CO disks of $1\farcm5$ radii (7 kpc) (figure 1), (2) the HI clouds can satisfy the criterion of ram pressure stripping while the molecular clouds cannot, and (3) the $\Sigma$---$f_{\rm mol}$ diagrams show unusually large $f_{\rm mol}$, indicating that the inner disks are highly HI deficient (figure 3). These three facts support the scenario that the ram pressure stripping of HI gas occurred even in the inner disks. 

On the other hand, NGC 4254 and NGC 4654 are located as far as $\sim 3 \arcdeg$ from the center of the cluster. They show the HI distribution extending to the outer disks and the $\Sigma$---$f_{\rm mol}$ relation being the same as that of field galaxies. Hence, most of the HI gas must have not yet been stripped in spite of the strongly distorted HI disks.

\subsubsection{Higher External Pressure due to the ICM-Pressure or Ram Pressure}
{Another possibility for making $f_{\rm mol}$ large is a higher external pressure due to the ICM-pressure, or the ram pressure, which might make the ISM pressure larger than that estimated by the surface density.

The ICM pressure affects all gas disks isotropically. The typical density and temperature of the ICM at the Virgo Cluster are $\rho \sim 10^{-4}$ cm$^{-3}$ and $T \sim 10^7$ K, respectively \citep{nul95}. On the other hand, the typical density and temperature of the interstellar molecular gas are $\rho \sim 10^{2}$ cm$^{-3}$ and $T \sim 10$ K, respectively. Since the pressure $P$ is proportional to $\rho T$, the ICM pressure and ISM pressure are comparable to each other near the cluster center.

The ram pressure would also increase the ISM pressure if the ISM were not stripped, as mentioned by \citet{ken89}. The typical velocity of a galaxy is $\sim 1000$ km s$^{-1}$ near the Virgo Cluster center. The velocity dispersion of a molecular cloud is typically $\sim 1$ km s$^{-1}$. Since the pressure $P$ is proportional to $\rho v^2$, the ram  and the ISM pressures are comparable near the cluster center. As a result, the ISM pressure would increase and the higher ISM pressure would make $f_{\rm mol}$ larger. 

We present the $\Sigma$---$f_{\rm mol}$ relation in figure \ref{fmol-Px2}, adopting $(P/P_0) = (\Sigma/\Sigma_0)^2+n^2$. 
The second term indicates an increment of the ISM pressure due to the ICM or the ram pressure. Figure \ref{fmol-Px2} shows that the ICM pressure makes $f_{\rm mol}$ much larger. 
Thus, we can conclude that a higher external pressure due to ICM or ram pressures could be the origin of the large $f_{\rm mol}$. }

\subsubsection{Larger Metallicity}
Because the metallicity strongly affects $f_{\rm mol}$, the larger metallicity might be an origin of the large $f_{\rm mol}$. 
In order to examine whether the large metallicity makes $f_{\rm mol}$ large, data on the metallicity distributions are necessary. However, there is no available data for these galaxies, except for NGC 4254. Therefore, we below present a new idea to calculate the metallicity distribution using the CO, HI, and H$\alpha$ data instead of using the metallicity data.

The metallicity is expressed using the oxygen abundance $(12+\log{[{\rm O}/{\rm H}]})$, which is correlated with the conversion factor $X$ \citep{ari96}. We here define 
\begin{equation}
\log{Z} = (12+\log{[{\rm O}/{\rm H}]}), 
\end{equation}
and adopt the relationship 
\begin{equation}
\log{\left({X\over 10^{20}}\right)} = -\log{Z}+9.3. 
\end{equation}
The metallicity of the solar neighborhood $\log{Z_0}$ is taken to be 8.9. 
We can calculate the H$_2$ surface density using $X$ if $Z$ is given. Combining the HI and H$_2$ surface densities, we obtain the total gas density $\Sigma$, which gives the pressure $P$ based on equation (\ref{eq-pressure}). 

The UV strength can be calculated using the H$\alpha$ data. H$\alpha$ images of the Virgo Cluster galaxies, except for NGC 4402, were archived by \citet{koo01}. Figure \ref{Halpha} shows contours of H$\alpha$ superimposed onto the CO images. We convolved the H$\alpha$ image with the same beams as the HI data. We adopted $U_0=2.42\times 10^{-7}$ erg cm$^{-2}$ s$^{-1}$ sr$^{-1}$ \citep{rey84}. For NGC 4402, we estimated $U$ while considering the Schmidt law,  
\begin{equation}
U = U_0\left({\Sigma \over \Sigma_0}\right)^n,  
\end{equation}
where we adopted $n=1.33$ \citep{kom05}.  

Thus, we can calculate $X$, $P$, and $U$ if $Z$ is given. Using obtained $P$, $U$, and $Z$, we can calculate the molecular fraction $f_{\rm mol}$ based on the model of \citet{elm93}. On the other hand, we can independently calculate $f_{\rm mol}$ using only the HI and H$_2$ surface densities. We search for an appropriate metallicity, which gives the same $f_{\rm mol}$ in the two ways. The searching range of the metallicity is adopted to be $6.68 < \log{Z} < 11.12$. 

This procedure is presented as a flow-chart in figure \ref{flowchart}. 

We plotted the calculated metallicity against the radius for each galaxy (figure \ref{metal}). Each point corresponds to each observed point. The metallicities of NGC 4254 and NGC 4654 show that $\log{Z}$ is about 9.5 -- 10 at the galactic center, and that it gradually declines. The metallicity of NGC 4254 was calculated by \citet{vil92} and \citet{zar94}. Our obtained values are almost consistent with their results. 

On the other hand, the metallicities $\log{Z}$ of NGC 4402, NGC 4569, and NGC 4579 often exceed 10. However, the former research shows that there is no galaxy that gives $\log{Z} > 10$.  
In addition, the dispersions of $\log{Z}$ of NGC 4402, NGC 4569, and NGC 4579 are  larger than the others. Moreover, the metallicities $\log{Z}$ of NGC 4402 and NGC 4569 increase with the radius. This tendency is unnatural because the metallicity $\log{Z}$ 
usually decreases with the radius \citep{vil92, zar94}. In the case of NGC 4579, there are many points where $Z$ 
cannot be calculated within the solution range $6.68 < \log{Z} < 11.12$ 
in the inner region. 

{Therefore, the possibility of higher metallicity is less plausible, because an abnormal metallicity distribution is necessary to reproduce such a large $f_{\rm mol}$. However, this possibility still cannot be ruled out, since nobody has directly measured the metallicities of NGC 4402, NGC 4569, and NGC 4579. }

\subsubsection{Lower UV Radiation Field}
{Because the UV strength affects $f_{\rm mol}$, a lower UV radiation field might be the origin of the large $f_{\rm mol}$. 
In order to examine this effect, we plotted the relation between the UV strength and the total gas density (figure \ref{gas-uv}), excluding the central 1 kpc to avoid H$\alpha$ emission from active galactic nuclei. This figure shows that there is little difference in the UV strength, $U/U_0$, for $\Sigma < 20 M_\odot $ pc$^{-2}$, although $U/U_0$ is lower in NGC 4569 than in NGC 4254 and NGC 4654 in the case of $\Sigma > 20 M_\odot $ pc$^{-2}$. 
Thus, it is difficult to attribute the large $f_{\rm mol}$ to a lower UV radiation field, because a large $f_{\rm mol}$ for $\Sigma < 20 M_\odot $ pc$^{-2}$ cannot be explained by a difference in the UV strength.}

\section{Summary}
We observed five Virgo spiral galaxies with the NRO 45 m telescope with BEARS in the $^{12}$CO$(J=1-0)$ line. Comparing the CO data with the HI data to investigate the environmental effect, we found that (1) the HI gas disks of NGC 4402, NGC 4569, and NGC 4579, located near the center of the cluster, are restricted within the inner disks, (2) the HI clouds can satisfy the criterion of the ram pressure stripping, while the molecular clouds cannot, and (3) the $\Sigma$--$f_{\rm mol}$ diagrams of the three galaxies show an unusually large $f_{\rm mol}$ { for a low total gas column density}. This large $f_{\rm mol}$ might imply that the HI gas is selectively stripped. 
On the other hand, NGC 4254 and NGC 4654, located far from the cluster center, have extended HI disks and show the normal $f_{\rm mol}$ expected for field galaxies. Most of the HI gas of these two galaxies must not have been stripped, in spite of the strongly distorted HI disks. 

A higher external pressure due to the ICM-pressure and/or ram pressure might be another possibility causing the unusually large $f_{\rm mol}$. 

Moreover, we examined whether differences in the intrinsic conditions (metallicity and UV) cause the unusually large $f_{\rm mol}$. As a result, we found that { an unusually high metallicity is necessary to explain such a large $f_{\rm mol}$, although it still cannot be ruled out.}

We also found that the large $f_{\rm mol}$ cannot be explained by only a difference in the UV radiation field. 

From these discussions, we note that the $\Sigma$---$f_{\rm mol}$ diagram is a good tool for investigating the environmental effect. An abnormal $f_{\rm mol}$ indicates that the ISM suffers cluster environmental effects: ram pressure stripping or higher external pressure . 

\acknowledgments
We are grateful to the members of the Nobeyama Radio Observatory. 
We appreciate that M. Honma kindly provided his code to calculate $f_{\rm mol}$. We would also like to thank T. Namba for his help with observations and reductions. 
F. E. is financially supported by a research fellowship from the Japan Society for the Promotion of Science for Young Scientists.

\begin{table}
\begin{center}
 \caption{Observational Parameter}
  \begin{tabular}{ccccccccccccccc}
   \hline\hline
   NGC &\multicolumn{3}{c}{R.A.(J2000)} &\multicolumn{3}{c}{Dec.(J2000)}
   &Morph.&B$_{\rm T}$&{\it i}& P.A.&$V_{\rm hel}$\\
   & h & m & s & d&m&s& &mag.&deg.&deg.&km s$^{-1}$\\
\hline
4254&12&18&49.61&+14&24&59.6&SA(s)c   &10.44&42&68&2405\\
4402&12&26&07.45&+13&06&44.7&Sb       &12.55&75&90&234\\
4569&12&36&49.82&+13&09&45.8&SAB(rs)ab&10.26&64&23&-235\\
4579&12&37&43.53&+11&49&05.5&SAB(rs)b &10.48&38&05&1520\\
4654&12&43&56.67&+13&07&36.1&SAB(rs)cd&11.10&52&125&1039\\ 
   \hline
   \hline
  \end{tabular}\\
Col. (1): Galaxy name. Cols. (2) and (3): Central position taken from \citet{sof03}. Col. (4): Morphological type from RC3 \citep{dev91}. Col. (5): Total B-band magnitude taken from RC3. Cols. (6) and (7): The inclination angle and the position angle (P.A.) taken from \citet{koo01} or \citet{pho93}. Col. (8): Systemic velocity taken from \citet{ken88} or RC3. 
\end{center}
\end{table}

\begin{table}
\begin{center}
 \caption{HI Data Parameter}
  \begin{tabular}{ccccc}
   \hline\hline
   NGC & Synth. Beam Size & $\Delta V$ & $\Delta T_{\rm B}$ & Reference\\
       & arcsec$^2$ &km s$^{-1}$ & K                  &           \\
\hline
4254   &$25\arcsec$ $\times$ $24\arcsec$ & 10.3       & 0.45               & 1\\ 
4402   &$21\arcsec$ $\times$ $17\arcsec$ & 25.0       & 1.7                & 2\\
4569   &$17\arcsec$ $\times$ $13\arcsec$ & 25.0       & 2.8                & 2\\
4579   &$19\arcsec$ $\times$ $18\arcsec$ & 25.0       & 2.7                & 2\\
4654   &$24\arcsec$ $\times$ $25\arcsec$ & 10.3       & 0.44               & 3\\ 
   \hline
   \hline
  \end{tabular}\\
Col.(1): Galaxy name. Col.(2): Synthesized beam size. Col.(3): Velocity resolution. Col.(4): R.m.s. noise in brightness temperature. Col.(5) References.---(1)\citet{pho93}; (2)\citet{cay90}; (3)\citet{pho95}
\end{center}
\end{table}

\begin{figure}
\epsscale{0.96}
\plotone{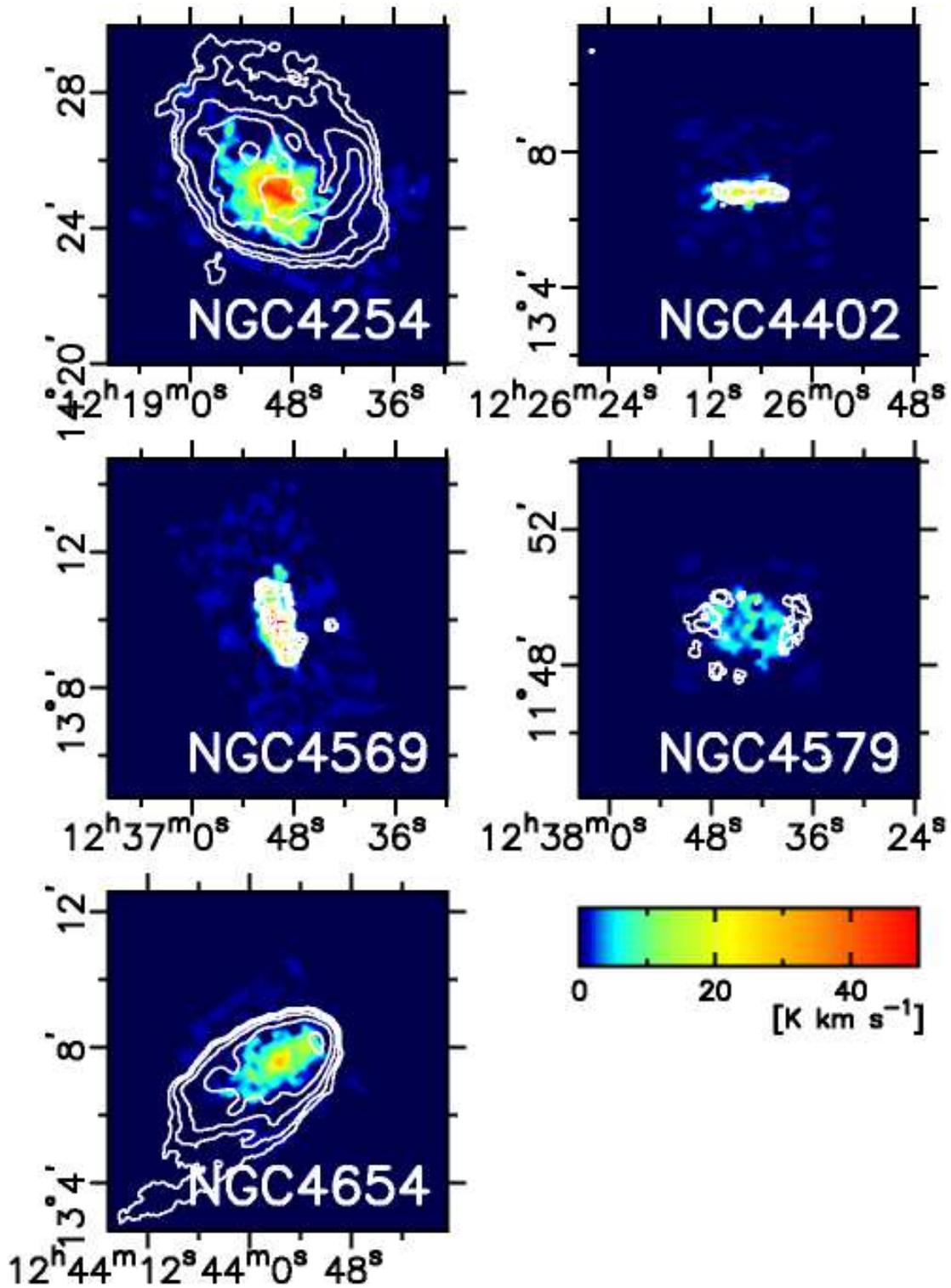}
\caption{Integrated CO intensity maps (color) and the integrated HI intensity maps (contour). The contour levels are 100, 200, 400, 800, 1600 K km s$^{-1}$.}
\end{figure}

\begin{figure}
\epsscale{0.6}
\plotone{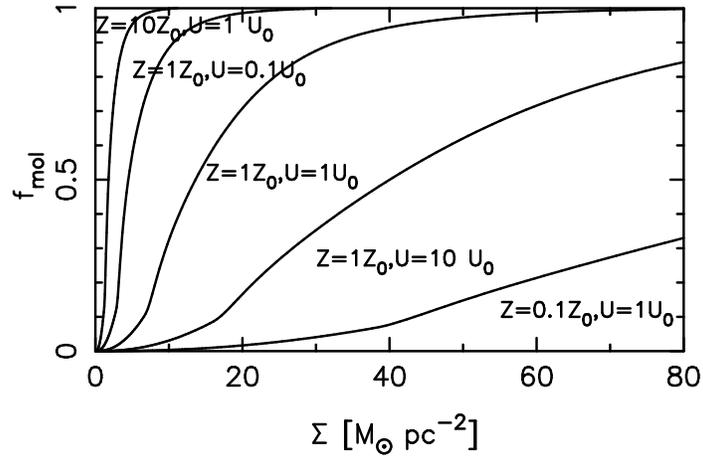}
\caption{Molecular fraction, $f_{\rm mol}$, against the surface density of the total gas $\Sigma$ calculated using a formula of \citet{elm93}. } 
\end{figure}

\begin{figure}
\epsscale{0.96}
\plotone{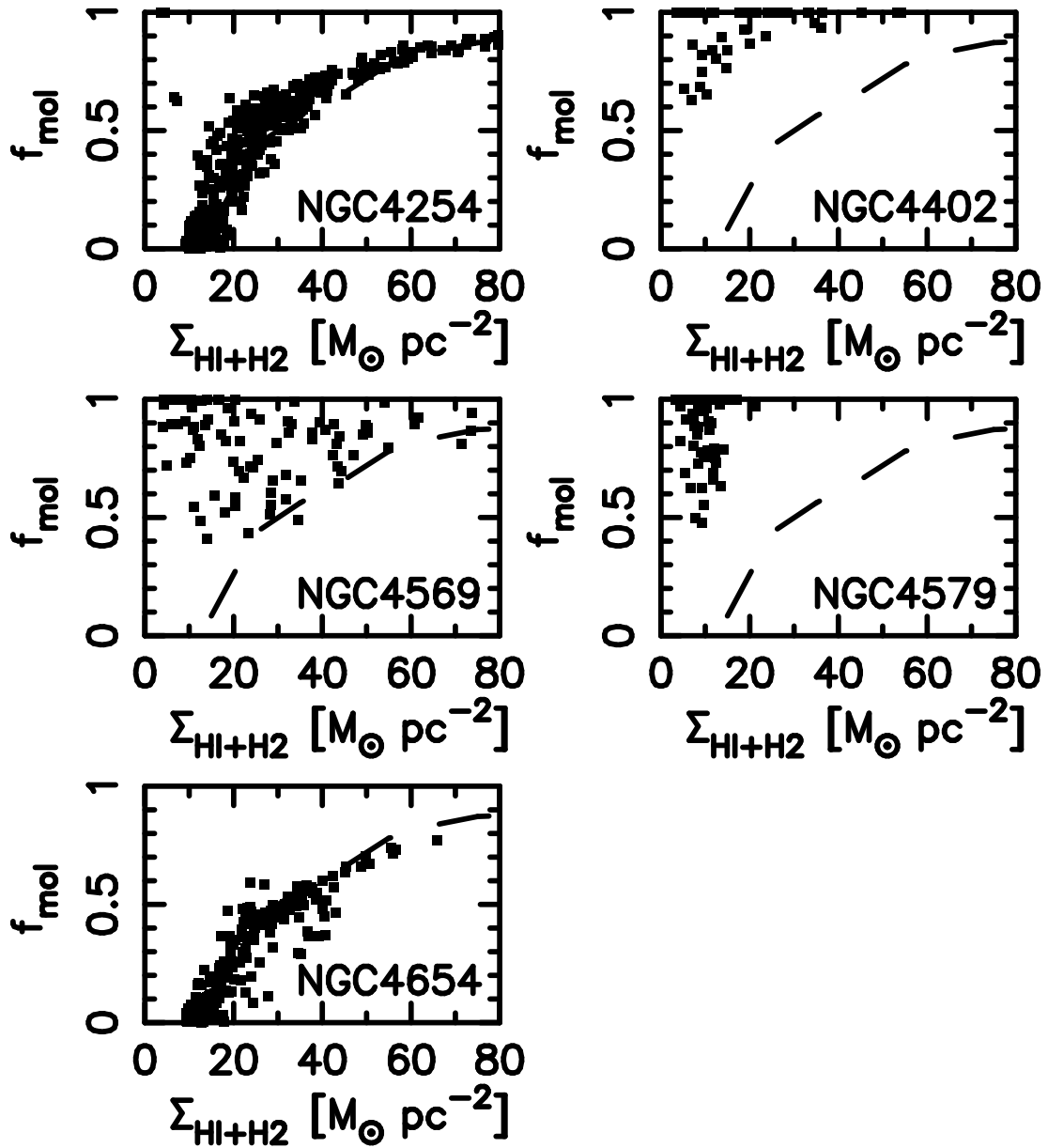}
\caption{Molecular fraction, $f_{\rm mol}$, against the surface density of the total gas $\Sigma$. The dashed curves in all panels denote the averaged $f_{\rm mol}$ of NGC 4254 and NGC 4654. 
The data of $\Sigma_{\rm HI} > 3.8 \times 10^{20}$ H cm$^{-2}$ or $\Sigma_{{\rm H}_2} > 7.0 \times 10^{19}$ H$_2$ cm$^{-2}$ (3$\sigma$) are plotted. }
\end{figure}

\begin{figure}
\epsscale{0.6}
\plotone{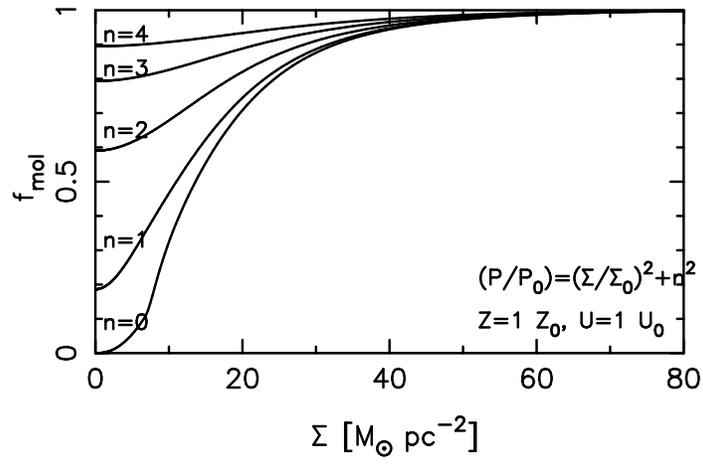}
\caption{Molecular fraction, $f_{\rm mol}$, against the surface density of the total gas $\Sigma$ calculated using a formula of \citet{elm93} and adopting $(P/P_0) = (\Sigma/\Sigma_0)^2 + n^2$ ($n=0$, 1, 2, 3, and 4). \label{fmol-Px2} }
\end{figure}

\begin{figure}
\epsscale{0.96}
\plotone{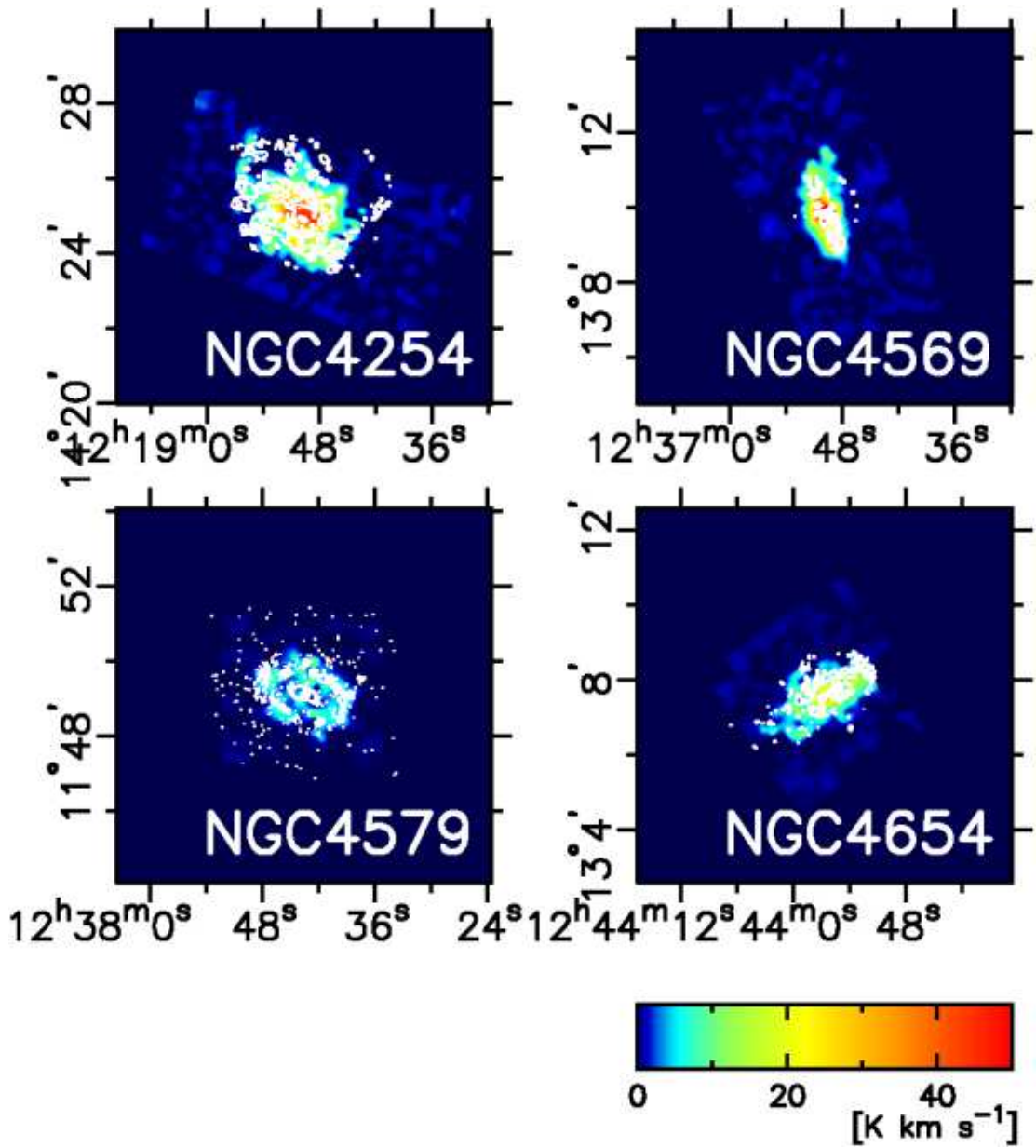}
\caption{Integrated CO intensity maps (color) and the H$\alpha$ maps (contour). The contour levels are $2.7\times 10^{-16}$ and $2.7\times 10^{-15}$ ergs cm$^{-2}$ s$^{-1}$ arcsec$^{-2}$. \label{Halpha}}
\end{figure}

\begin{figure}
\epsscale{0.96}
\plotone{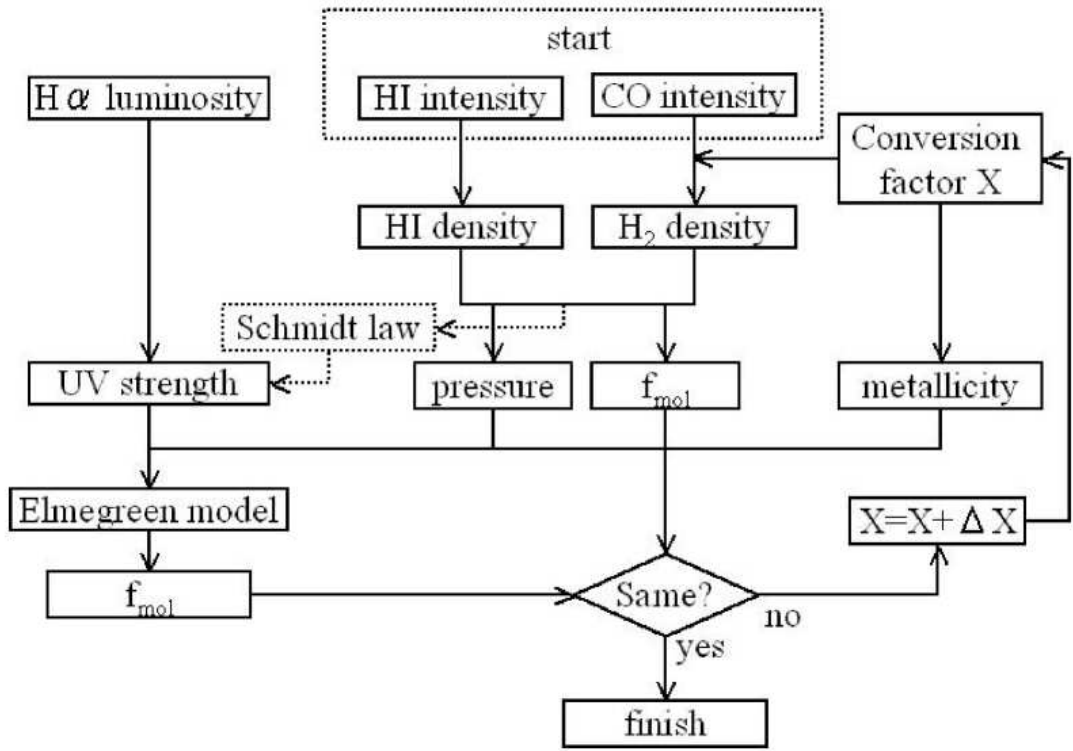}
\caption{Flow chart to calculate the metallicity using HI, CO, and H$_\alpha$ data. \label{flowchart} }
\end{figure}

\begin{figure}
\epsscale{0.96}
\plotone{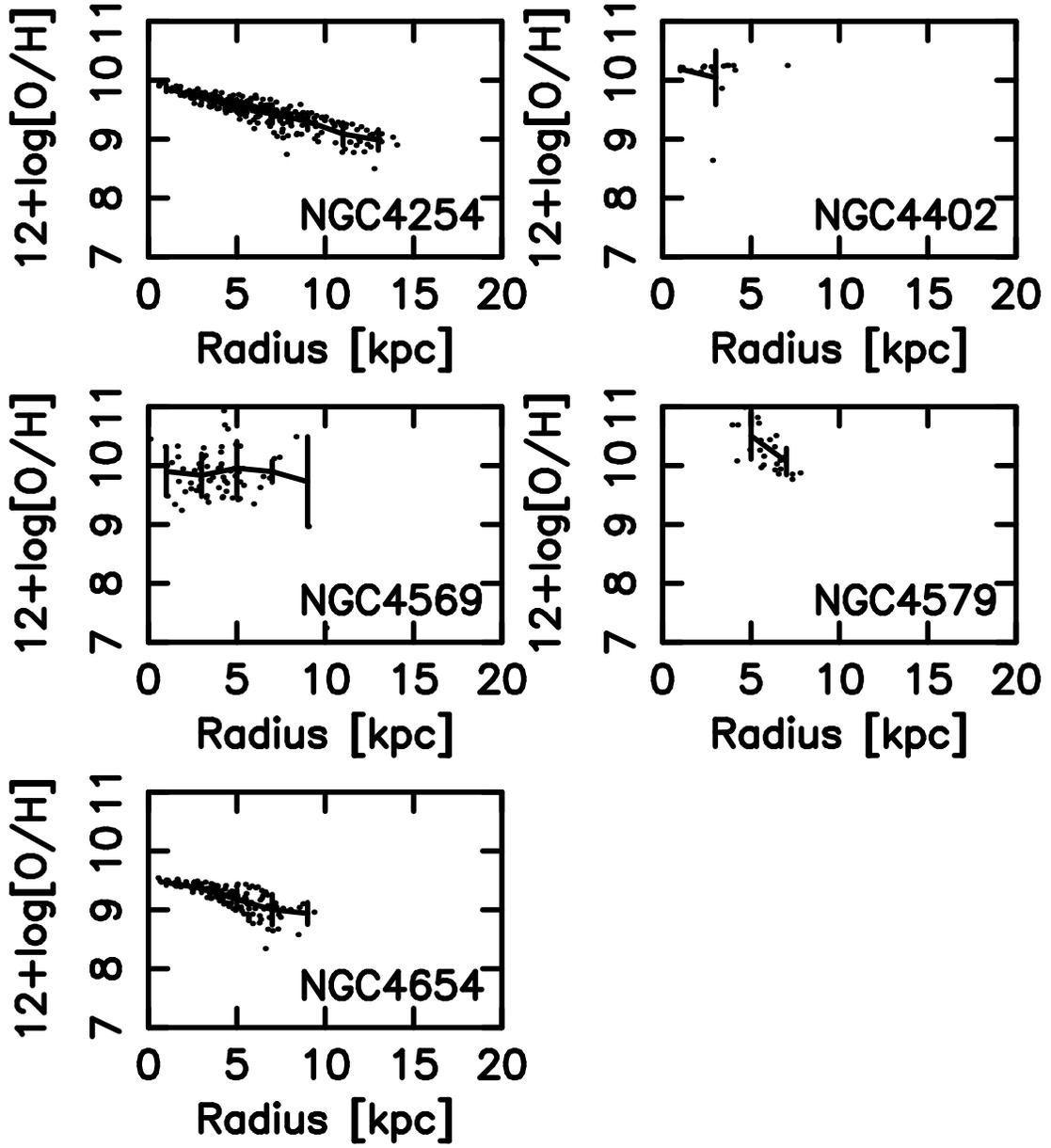}
\caption{Metallicity $\log{Z}=(12+\log{[{\rm O}/{\rm H}]})$ against the radius. The metallicities were calculated using the data of $\Sigma_{\rm HI} > 3.8 \times 10^{20}$ H cm$^{-2}$ or $\Sigma_{{\rm H}_2} > 7.0 \times 10^{19}$ H$_2$ cm$^{-2}$ (3$\sigma$). The solid curves and vertical segments indicate the mean values and standard deviations at each radius. \label{metal} }
\end{figure}

\begin{figure}
\epsscale{0.96}
\plotone{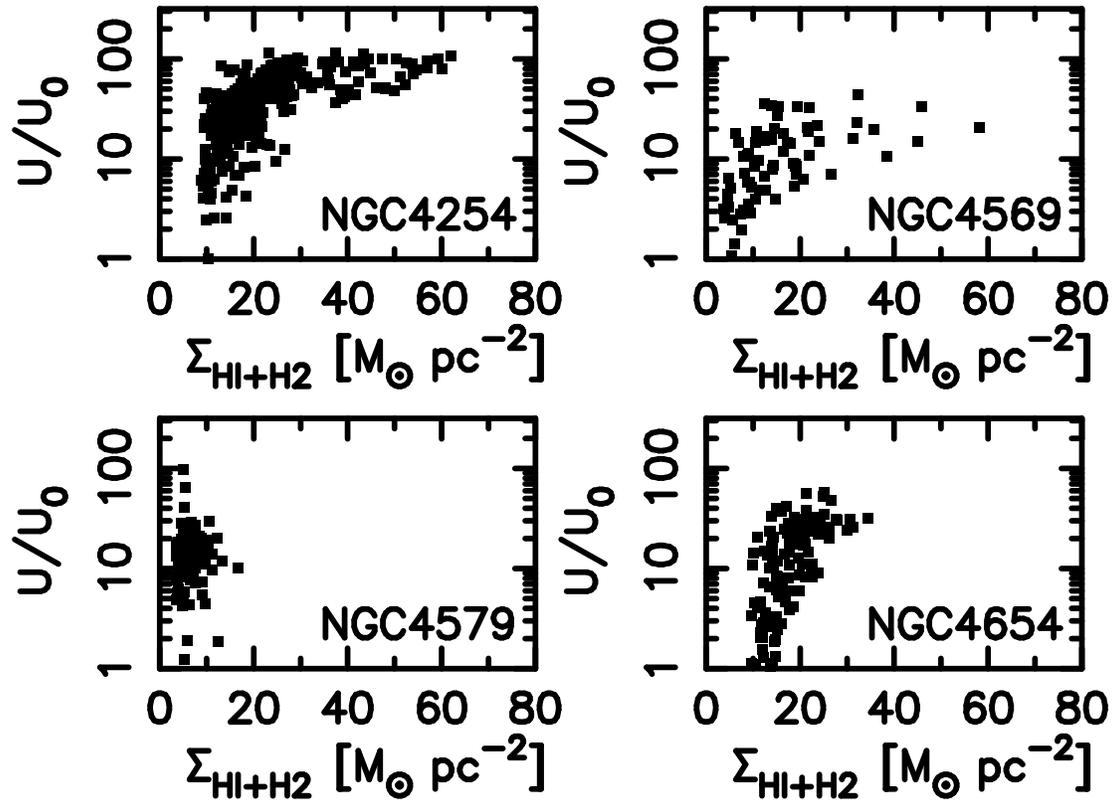}
\caption{Relationship between the UV strength $(U/U_0)$ and surface density of the total gas $\Sigma$. \label{gas-uv} }
\end{figure}

\end{document}